\documentclass[runningheads]{svmult}
\usepackage{makeidx}   % allows index generation
\usepackage{graphicx}  % standard LaTeX graphics tool
\usepackage{subeqnar}  % subnumbers individual equations
\usepackage{multicol}  % used for the two-column index
\usepackage{cropmark} % cropmarks for pages without
\usepackage{physprbb}  % centered layout of diverse elements, etc.
\begin{document}
\title*{Neutralino Dark Matter vs Galaxy Formation}
\toctitle{Neutralino Dark Matter vs Galaxy Formation}
% allows explicit linebreak for the table of content
%
%
\titlerunning{Neutralino Dark Matter vs Galaxy Formation}
% allows abbreviation of title, if the full title is too long
% to fit in the running head
%
\author{Paolo Gondolo\inst{1}}
\authorrunning{Paolo Gondolo}
% if there are more than two authors,
% please abbreviate author list for running head
%
%
\institute{Max Planck Institute for Physics, F\"ohringer Ring 6, 80805 Munich,
 Germany}

\maketitle              % typesets the title of the contribution

\begin{abstract}
  Neutralino dark matter may be incompatible with current cold dark matter
  models with cuspy dark halos, because excessive synchrotron radiation may
  originate from neutralino annihilations close to the black hole at the
  galactic center.
\end{abstract}

We report results obtained in~\cite{g00}, to which we refer for further
details.

The composition of dark matter is one of the major issues in cosmology. A
popular candidate for non-baryonic cold dark matter is the lightest neutralino
appearing in a large class of supersymmetric models~\cite{jun96}.
In a wide range of supersymmetric parameter space, relic neutralinos from the
Big Bang are in principle abundant enough to account for the dark matter in our
galactic halo~\cite{eds97}.

A generic prediction of cold dark matter models is that dark matter halos
should have steep central cusps, meaning that their density rises as
$r^{-\gamma}$ to the center.  Semi-analytical calculations find a cusp slope
$\gamma$ between $\sim 1$~\cite{sw98} and 2~\cite{hs85}. Simulations find a
slope $\gamma$ ranging from 0.3~\cite{kra98} to 1~\cite{nfw} to
1.5~\cite{fuk97}.  It is unclear if dark matter profiles in real galaxies and
galaxy clusters have a central cusp or a constant density core.

There is mounting evidence that the non-thermal radio source Sgr A$^*$ at the
galactic center is a black hole of mass $ M \sim 3 \times 10^6 \, M_{\odot}$.
This inference is based on the large proper motion of nearby
stars~\cite{ghe98}, the spectrum of Sgr A$^*$ (e.g.~\cite{mez96,nar98}), and
its low proper motion~\cite{bac99}. It is difficult to explain these data
without a black hole~\cite{mao98}.

\begin{figure}[b]
\begin{center}
\includegraphics[width=.9\textwidth]{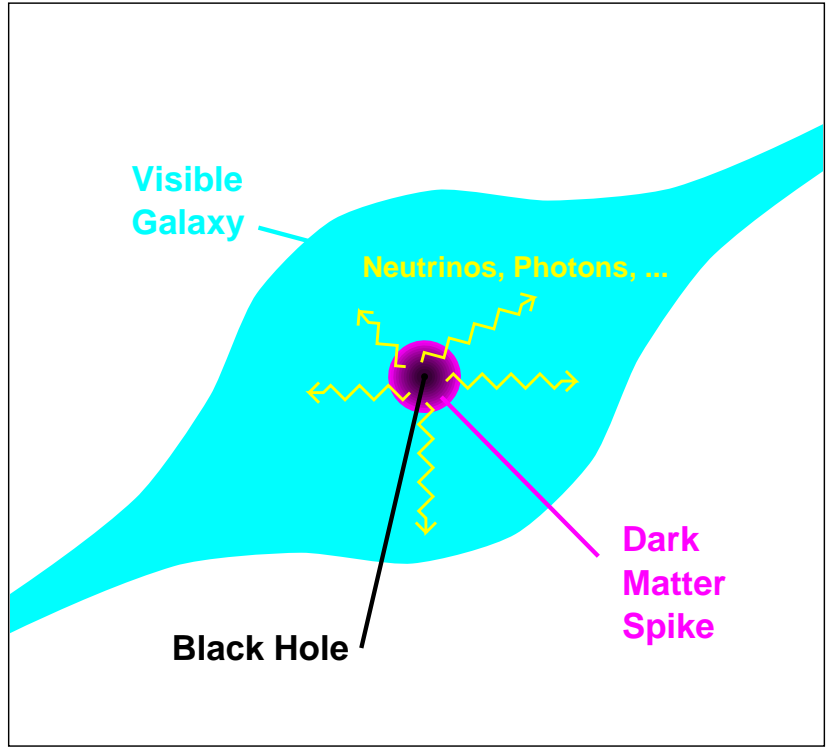}
\end{center}
\caption[]{Artist's impression of neutralino dark matter annihilations into
  neutrinos, photons, and other standard particles in the very dense dark
  matter spike which may form around the black hole at the galactic center.}
\label{eps1.1}
\end{figure}

The black hole at the galactic center modifies the distribution of dark matter
in its surroundings~\cite{gs99}, creating a high density dark matter region
called the spike -- to distinguish it from the above mentioned cusp (see
Fig.~\ref{eps1.1} for an illustration). Signals from particle dark matter
annihilation in the spike may be used to discriminate between a central cusp
and a central core.  With a central cusp, the annihilation signals from the
galactic center increase by many orders of magnitude. With a central core, the
annihilation signals do not increase significantly.

Stellar winds are observed to pervade the inner parsec of the
galaxy~\cite{mez96}, and are supposed to feed the central black hole
(e.g.~\cite{nar98,cok99}). These winds carry a magnetic field whose measured
intensity is a few milligauss at a distance of $\sim 5{\rm pc}$ from the
galactic center~\cite{yus96}. The magnetic field intensity can rise to a few
kilogauss at the Schwarzschild radius of the black hole in some accretion
models for Sgr A$^*$~\cite{mel92}. The existence and strength of a magnetic
field in the inner parsec of the galaxy is crucial to our argument.

In~\cite{g00} we examine the radio emission from neutralino dark matter
annihilation in the central spike. (Previous studies of radio emission from
neutralino annihilation at the galactic center have considered an $r^{-1.8}$
cusp but no spike~\cite{ber94}.)  Radio emission is due to synchrotron
radiation from annihilation electrons and positrons in the magnetic field
around Sgr A$^*$. Comparing the radio emission from the neutralino spike with
the measured Sgr A$^*$ spectrum, we find that neutralino dark matter in the
minimal supersymmetric standard model is incompatible with a dark matter cusp
extending to the galactic center.

Since the strength and structure of the magnetic field around Sgr A$^*$ is only
known to some extent (see discussion in~\cite{g00}), we consider three
simple but relevant models for the magnetic field and the electron/positron
propagation.  In model A, we assume that the magnetic field is uniform across
the spike, with strength $B=1{\rm mG}$, and that the electrons and positrons
lose all their energy into synchrotron radiation without moving significantly
from their production point.  In model B, we also assume that the magnetic
field is uniform across the spike with strength $B=1{\rm mG}$, but that the
electrons and positrons diffuse efficiently and are redistributed according to
a gaussian encompassing the spike (we take the gaussian width $\lambda=1$ pc).
In model C, we assume that the magnetic field follows the equipartition value
$B = 1 \mu{\rm G} (r/{\rm pc})^{-5/4}$ (from ref.~\cite{mel92}) and that the
electrons and positrons lose all their energy into synchrotron radiation
without moving significantly from their production point. In addition, in model
C, we neglect synchrotron self-absorption.

Under these assumptions, we obtain the following results.

\begin{figure}[b]
\begin{center}
\includegraphics[width=.9\textwidth]{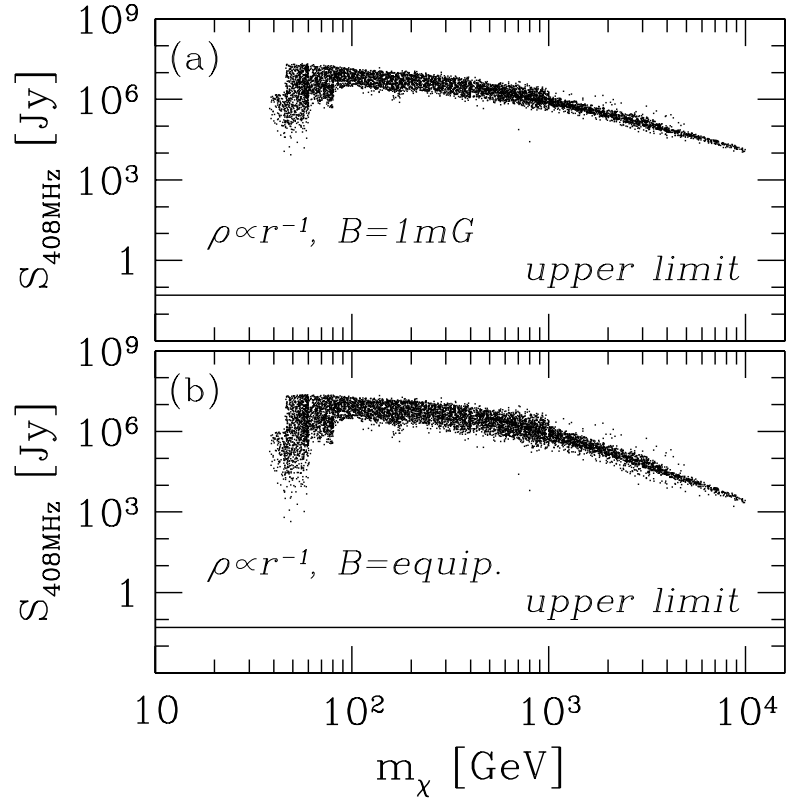}
\end{center}
\caption[]{Expected radio emission from the galactic center at 408 MHz from
  neutralino annihilations in the dark matter spike, assuming a
  Navarro-Frenk-White profile and (a) a uniform magnetic field of 1 mG, (b) a
  magnetic field at the equipartition value. All models exceed the present
  upper bound by several orders of magnitude.}
\label{eps1.2}
\end{figure}

If a dark matter cusp extends to the galactic center, the neutralino cannot be
the dark matter in our galaxy. For example, let us assume that the halo profile
is of the Navarro-Frenk-White form~\cite{nfw}, namely $\rho \propto r^{-1}$ in
the central region. Fig.~\ref{eps1.2} shows the expected radio fluxes $S_\nu =
L_\nu/4\pi D^2$ at 408 MHz and the upper limit from~\cite{dav76}. The upper
panel is for model A, the lower panel for model C.  Results of model B are
similar to those of model A. Irrespective of the assumption on the magnetic
field or the $e^\pm$ propagation, all points in supersymmetric parameter space
where the neutralino would be a good dark matter candidate are excluded by
several orders of magnitude.

\begin{figure}[b]
\begin{center}
\includegraphics[width=.9\textwidth]{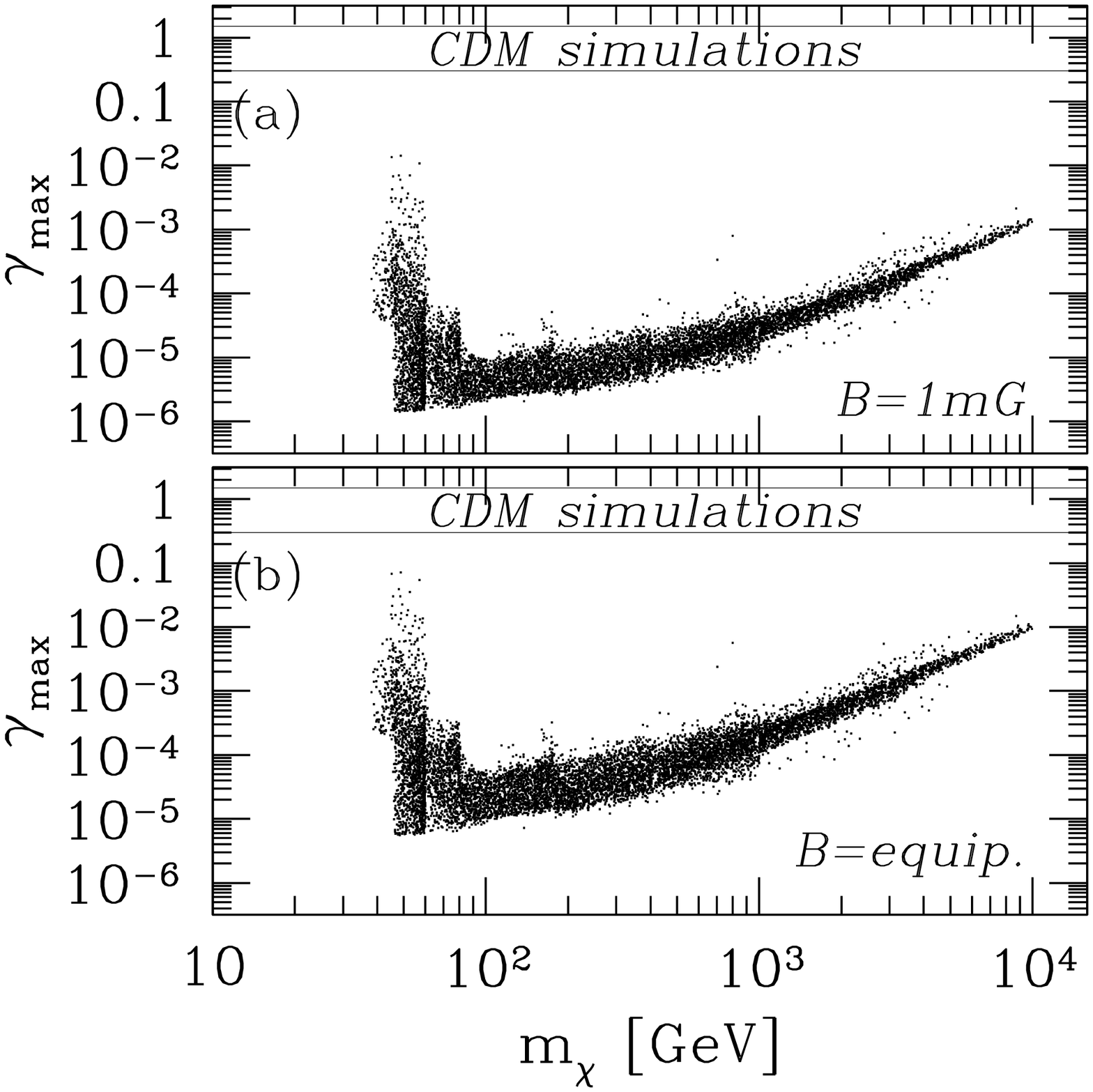}
\end{center}
\caption[]{Upper bound on the inner halo slope $\gamma$ imposed by the
  constraint on the radio emission from the galactic center at 408 MHz,
  assuming (a) a uniform magnetic field of 1 mG, and (b) a magnetic field at
  the equipartition value. Each dot corresponds to a point in supersymmetric
  parameter space.  The results of cold dark matter simulations are much higher
  than the upper bounds.}
\label{eps1.3}
\end{figure}

Conversely, if the neutralino is the dark matter, there is no steep dark matter
cusp extending to the galactic center. We see this by lowering the cusp slope
$\gamma$ until the expected flux at 408 MHz decreases below the upper limit.
We obtain a different maximum value $\gamma_{\rm max}$ for each point in
supersymmetric parameter space. These values are plotted in Fig.~\ref{eps1.3}
together with the range $0.3 \stackrel{<}{{}_\sim} \gamma \stackrel{<}{{}_\sim}
1.5$ obtained in cold dark matter simulations. The upper bounds $\gamma_{\rm
  max}$ are generally orders of magnitude smaller than the simulation results.

We conclude that neutralino dark matter in the minimal supersymmetric standard
model is incompatible with a dark matter cusp extending to the galactic center.
If there is a dark matter cusp extending to the center, we can exclude the
neutralino in the minimal supersymmetric standard model as a dark matter
candidate.  Conversely, if the dark matter of the galactic halo is the lightest
neutralino in the minimal supersymmetric standard model, we can exclude that a
dark matter cusp extends to the center of the galaxy. Our conclusions are based
on the presence of a magnetic field in the central parsec of our galaxy: if the
magnetic field would be absent, there would be no synchrotron emission from
annihilation electrons and positrons, and hence no synchrotron limits on
neutralino dark matter and cuspy dark halos.

\begin{thebibliography}{8.}
\addcontentsline{toc}{section}{References}

\bibitem{g00} P. Gondolo: `Either Neutralino Dark Matter or Cuspy Dark
  Halos,' preprint hep-ph/0002226

\bibitem{jun96} G. Jungman, M. Kamionkowski, and K. Griest, Phys.\ Rep.\ {\bf
    267}, 195 (1996)

\bibitem{eds97} J. Edsj\"o and P. Gondolo, Phys.\ Rev.\ {\bf D56}, 1879 (1997)
  
\bibitem{sw98} D. Syer and S.D.M. White, MNRAS {\bf 293}, 337 (1998)

\bibitem{hs85} Y. Hoffman and J. Shaham, Ap.\ J. {\bf 297}, 16 (1985)

\bibitem{kra98} A.V. Kravtsov, A.A. Klypin, J.S. Bullock, and J.R. Primack,
  Ap.\ J. {\bf 502}, 48 (1998); J. S. Bullock et al., astro-ph/9908159

\bibitem{nfw} J.F. Navarro, C.S. Frenk, and S.D.M. White, Ap.\ J. {\bf 462},
  563 (1996); ibid.\ {\bf 490}, 493 (1997)
  
\bibitem{fuk97} T. Fukushige and J. Makino, Ap.\ J. {\bf 477}, L9 (1997); B.
  Moore, F. Governato, T. Quinn, J. Stadel, and G. Lake, Ap.\ J. {\bf 499}, L5
  (1998); S. Ghigna et al., astro-ph/9910166
  
\bibitem{ghe98} A. Eckart and R. Genzel, Nature {\bf 383}, 415 (1996); MNRAS

  {\bf 284}, 576 (1997); A.L. Ghez, B.M. Klein, M. Morris, and E.E. Becklin,
  Ap.\ J.  {\bf 509}, 678 (1998); R. Genzel, C. Pichon, A. Eckart,
  O.E. Gerhard, and T. Ott, astro-ph/0001428
  
\bibitem{mez96} P.G. Mezger, W.J. Duschl, and R. Zylka, Astron.\ Astrophys.\ 
  Rev.\ {\bf 7}, 289 (1996); F. Yusef-Zadeh, F. Melia, and M.
  Wardle, Science {\bf 287}, 85 (2000)

\bibitem{nar98} R. Narayan, R. Mahadevan, J.E. Grindlay, R.G. Popham, and
  C. Gammie, Ap.\ J. {\bf 492}, 554 (1998)

\bibitem{bac99} D.C. Backer and R.A. Sramek, Ap.\ J. {\bf 524}, 805 (1999)

\bibitem{mao98} E. Maoz, Ap.\ J. {\bf 494}, L131 (1998)

\bibitem{gs99} P. Gondolo and J. Silk, Phys.\ Rev.\ Lett.\ {\bf 83}, 1719
  (1999)
  
\bibitem{cok99} R. Coker, F. Melia, and H. Falcke, Ap.\ J. {\bf 523}, 642
  (1999)

\bibitem{yus96} F. Yusef-Zadeh, D.A. Roberts, W.M. Goss, D. Frail, and
  A. Green, Ap.\ J. {\bf 466}, L25 (1996); Ap.\ J. {\bf 512}, 230 (1999)

\bibitem{mel92} F. Melia, Ap.\ J. {\bf 387}, L25 (1992)

\bibitem{ber94} V. Berezinsky, A. Bottino, and G. Mignola, Phys.\ Lett.\ {\bf
    B325}, 136 (1994)

\bibitem{dav76} R.D. Davies, D. Walsh, and R. Booth, MNRAS {\bf 177}, 319
  (1976)

\end{thebibliography}
\end{document}